\documentclass[aps,pre,reprint,groupedaddress,showpacs,floatfix]{revtex4-1}

\usepackage{amsmath}
\usepackage{amsfonts}
\usepackage{color}
\usepackage{graphicx}
\usepackage{times}
\usepackage{ifpdf}

\begin{document}

\title{Simplifying the complexity of pipe flow}

\author{Dwight Barkley}
\email[Email: ]{D.Barkley@warwick.ac.uk} 
\affiliation{Mathematics Institute,
University of Warwick, 
Coventry CV4 7AL, United Kingdom \\ 
PMMH (UMR 7636 CNRS - ESPCI - Univ Paris 06 - Univ Paris 07), 
10 rue Vauquelin, 75005 Paris France 
}

\date{\today}

\begin{abstract}

Transitional pipe flow is modeled as a one-dimensional excitable and bistable
medium.  Models are presented in two variables, turbulence intensity and mean
shear, that evolve according to established properties of transitional
turbulence. A continuous model captures the essence of the puff-slug
transition as a change from excitability to bistability. A discrete model,
that additionally incorporates turbulence locally as a chaotic repeller,
reproduces almost all large-scale features of transitional pipe flow. In
particular it captures metastable localized puffs, puff splitting, slugs, a
continuous transition to sustained turbulence via spatiotemporal
intermittency (directed percolation), 
and a subsequent increase in turbulence fraction towards uniform, 
featureless turbulence.

\end{abstract}




\maketitle


The transition to turbulence in pipe flow has been the subject of study for
over 100 years~\cite{Reynolds:1883}, both because of its
fundamental role in fluid mechanics and because of the detrimental
consequences of turbulent transition in many practical situations.
There are at least two features of the problem that make it fascinating, but
also difficult to analyze. The first is that when turbulence appears, it
appears abruptly~\cite{Reynolds:1883}, and not through a sequence of
transitions each increasing the dynamical complexity of the flow.
Turbulence is triggered by finite-sized disturbances to linearly stable
laminar flow~\cite{Darbyshire:1995,Hof:2003p846,Peixinho:2007}.
This hysteretic, or subcritical, aspect of the problem limits the applicability
of linear and weakly nonlinear theories.
The second complicating feature is the intermittent form turbulence takes in
the transitional regime near the minimum Reynolds number (non-dimensional flow
rate) for which turbulence is observed~\cite{Reynolds:1883, Rotta:1956, 
Wygnanski:1973, Moxey:2010}.
In sufficiently long pipes, localized patches of turbulence (puffs) may
persist for extremely long times before abruptly reverting to laminar flow
\cite{Faisst:2004p898, Peixinho:2006, Hof:2006p607, 
Willis:2007p833, Schneider:2008p642, Hof:2008p914, Avila:2010p839,
Kuik:2010p911}. In other cases, turbulent patches may spread by contaminating
nearby laminar flow (puff splitting and slugs)~\cite{Wygnanski:1973,
Wygnanski:1975, Nishi:2008, Moxey:2010, Duguet:2010p2, Avila:2011}.
While minimal models have been very useful in understanding generic features
of intermittency in subcritical shear flows \cite{Pomeau:1986, Chate:1988p8,
Chate:1988p48, Bottin:1998p305, Manneville:2009p316}, such models do not
capture the puffs, puff splitting, and slugs that are essential to the
character of pipe flow.
In this paper I argue that transitional pipe flow should be viewed in the
context of excitable and bistable media. With this perspective I present
models, based on the interaction between turbulence and the mean shear, that
both capture and organize most large-scale features of transitional pipe flow.

\begin{figure}
\centering
\includegraphics[width=3.375in]{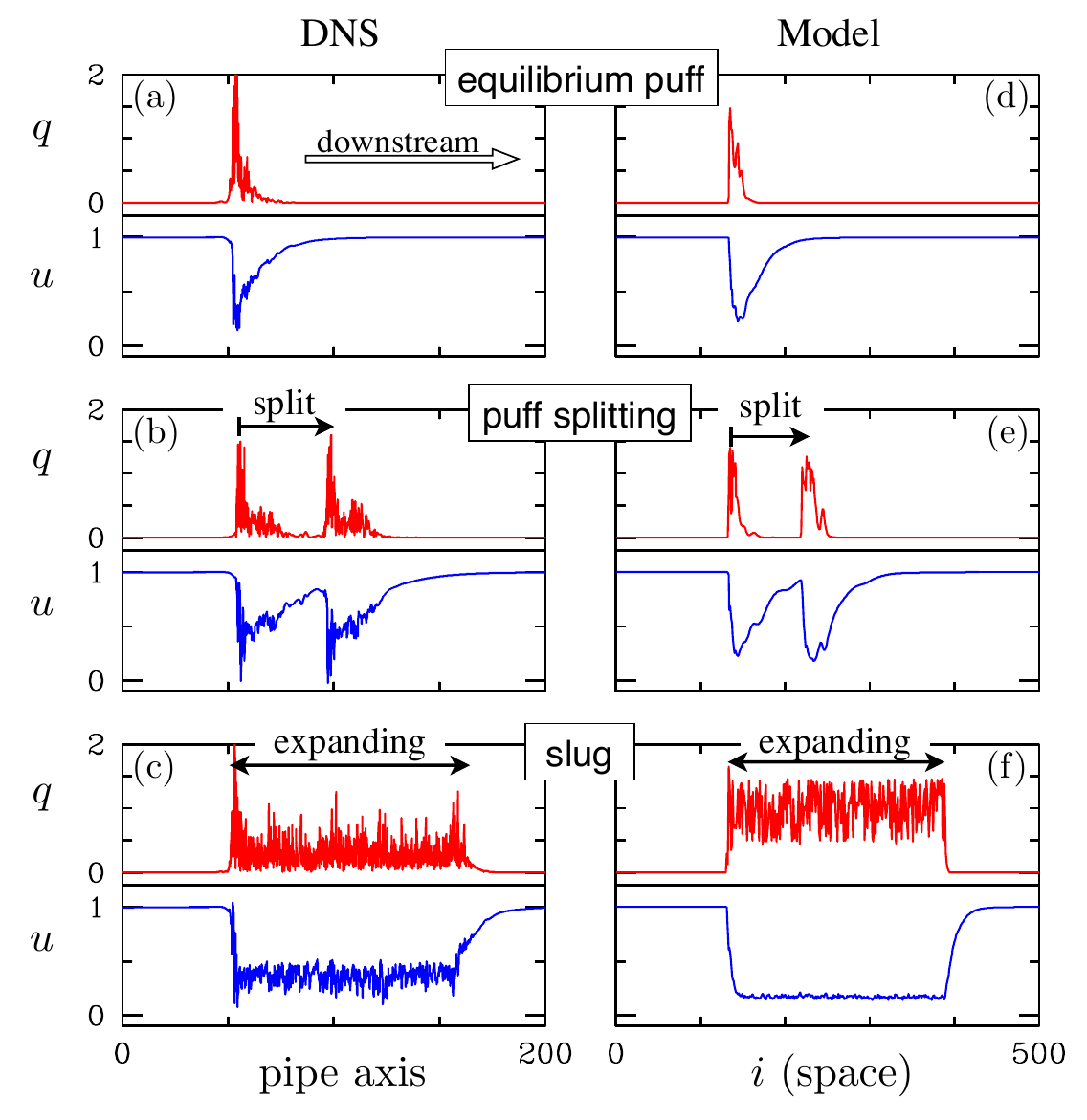}
\caption
{(Color online) Regimes of transitional pipe flow. Left column is from full
DNS with $4 \times 10^7$ degrees of freedom in a periodic pipe 200D long. Flow
is from left to right. Shown are instantaneous values of turbulence intensity
$q$ and axial velocity $u$ along the pipe axis.
(a) Equilibrium puff at $Re = 2000$.  (b) Puff splitting at $Re = 2275$. The
downstream (right) puff split from the upstream one at an earlier time. (c)
Expanding slug flow at $Re = 3200$.
Right column shows corresponding states from the simple one-dimensional
model~\eqref{eq:q_map}-\eqref{eq:alpha} (d) $R = 2000$, (e) $R=2100$, and (f)
$R=3200$.  }
\label{fig:DNS_model}
\end{figure}

Figure~\ref{fig:DNS_model} summarizes the three important dynamical regimes of
transitional pipe flow.  The left column shows results from direct numerical
simulations (DNS)~\cite{Moxey:2010,Blackburn:2004p1087}.
Quantities are nondimensionalized by the pipe diameter $D$ and the mean (bulk)
velocity $U$.  The Reynolds numbers is $Re = DU/\nu$, where $\nu$ is kinematic
viscosity.
Flows are well represented by two quantities, the turbulence intensity $q$ and
the axial (streamwise) velocity $u$, sampled on the pipe axis.
Specifically, $q$ is the magnitude of transverse fluid velocity (scaled up by
a factor of 6).  The centerline velocity $u$ is relative to the mean velocity
and is a proxy for the state of the mean shear that conveniently lies between
0 and 1.
At low $Re$, as in Fig.~\ref{fig:DNS_model}(a), turbulence occurs in localized
patches propagating downstream with nearly constant shape and speed. These
are called equilibrium puffs~\cite{Wygnanski:1975, Darbyshire:1995,
Nishi:2008}, a misnomer since at low $Re$ puffs are only metastable and
eventually revert to laminar
flow, i.e.\ decay~\cite{Faisst:2004p898, Peixinho:2006, Hof:2006p607,
Willis:2007p833, Schneider:2008p642, Hof:2008p914, Avila:2010p839,
Kuik:2010p911}. Asymptotically the flow will be laminar parabolic flow, 
$(q=0, u=1)$, throughout the pipe.
For intermediate $Re$, as in Fig.~\ref{fig:DNS_model}(b), puff splitting
frequently occurs~\cite{Wygnanski:1975, Nishi:2008, Moxey:2010, Avila:2011}.
New puffs are spontaneously generated downstream from existing ones and the
resulting pairs move downstream with approximately fixed separation.  Further
splittings will occur and interactions will lead asymptotically to a highly
intermittent mixture of turbulent and laminar flow~\cite{Rotta:1956,
Moxey:2010}.
At yet higher $Re$, turbulence is no longer confined to localized patches, but
spreads aggressively in so-called slug flow~\cite{Wygnanski:1973, Nishi:2008},
as illustrated in Fig.~\ref{fig:DNS_model}(c). The asymptotic state is uniform,
featureless turbulence throughout the pipe~\cite{Moxey:2010}.


Models for these dynamics will be based on the following known physical
features.
At the upstream (left in Fig.~\ref{fig:DNS_model}) edge of turbulent patches,
laminar flow abruptly becomes turbulent.
Energy from the laminar shear is rapidly converted into turbulent motion and
this results in a rapid change to the mean shear profile~\cite{Wygnanski:1973,
Hof:2010p65}.
In the case of puffs, the turbulent profile is not able to sustain turbulence
and thus there is a reverse transition~\cite{Wygnanski:1973, Sreenivasan:1979}
from turbulent to laminar flow on the downstream side of a puff. In the case
of slugs, the turbulent shear profile can sustain turbulence indefinitely;
there is no reverse transition and slugs grow to arbitrary streamwise
length~\cite{Wygnanski:1973,Nishi:2008}.
On the downstream side of turbulent patches the mean shear profile recovers
slowly~\cite{Sreenivasan:1979}, seen in the behavior of $u$ in Fig. 1.
Crucially, the degree of recovery dictates how susceptible the flow is to
re-excitation into turbulence~\cite{Hof:2010p65}.

These are the characteristics of excitable and bistable
media~\cite{TYSON:1988p1143, KeenerSneyd}.  
In fact the puff in Fig.~\ref{fig:DNS_model}(a) bears a close resemblance
to an action potential in a nerve axon~\cite{HODGKIN:1952p1148}.
Linearly stable parabolic flow is the excitable rest state, turbulence is the
excited state, and the mean shear is the recovery variable controlling the
threshold for excitation. Thus, I propose to model pipe flow as a generic
excitable and bistable medium incorporating the minimum requisite features of
pipe turbulence.  The models are expressed in variables $q$ and $u$ depending
on distance along the pipe.

Consider first the continuous model
\begin{eqnarray}
q_t + U q_x & = & q \left( u + r - 1 - (r + \delta) (q -1)^2  \right) 
    + q_{xx}, \label{eq:q_pde} \\
u_t + U u_x & = & \epsilon_1 (1 - u) - \epsilon_2 u q - u_x, \label{eq:u_pde} 
\end{eqnarray}
where $r$ plays the role of $Re$. $U$ accounts for downstream advection by the
mean velocity, and is otherwise dynamically irrelevant since it can
be removed by a change of reference frame. The model includes minimum
derivatives, $q_{xx}$ and $u_x$, needed for turbulent regions to excite
adjacent laminar ones and for left-right symmetry breaking.

\begin{figure}
\centering
\includegraphics[width=3.2in,clip]{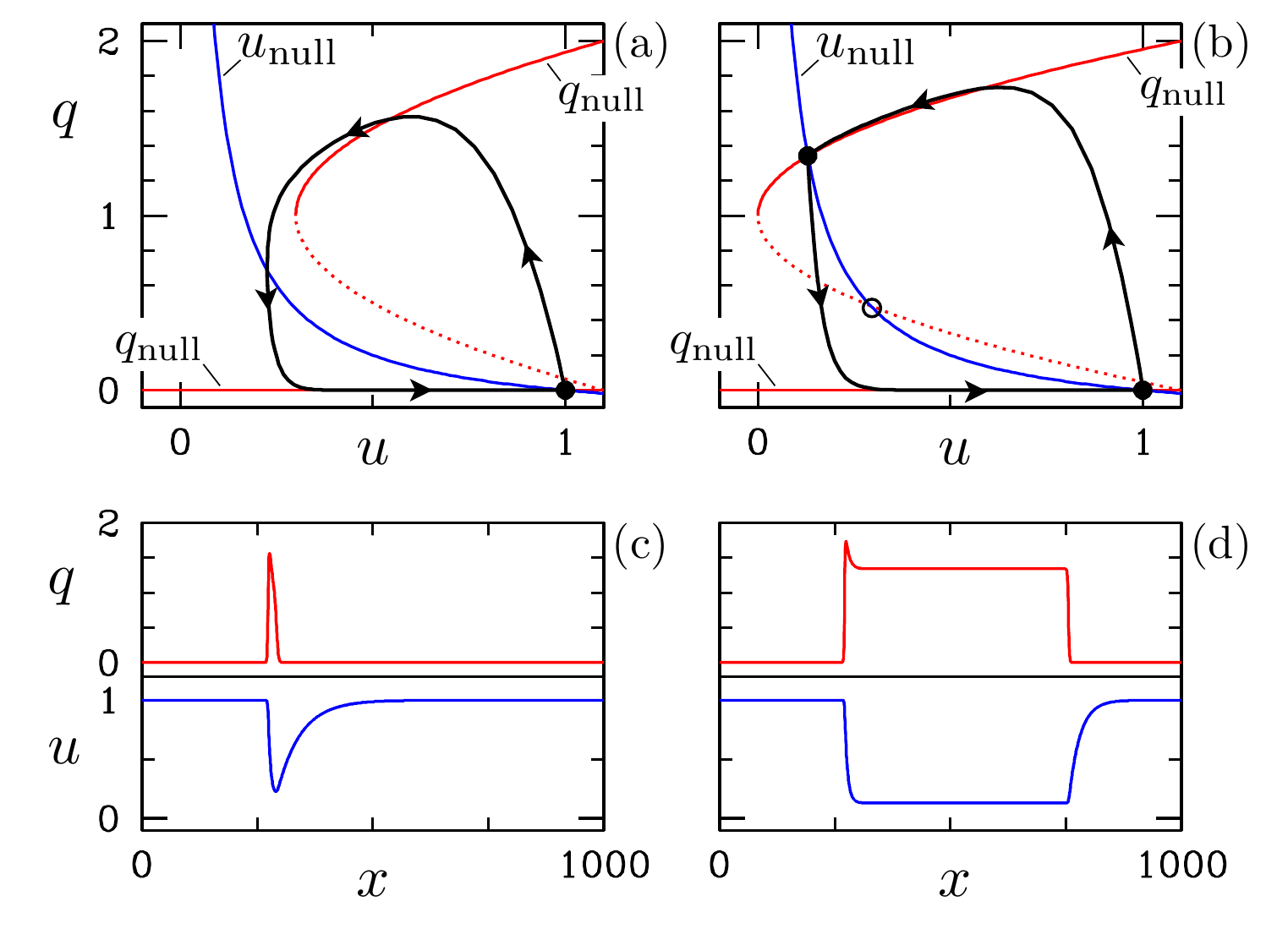}
\caption {(Color online) 
The distinction between puffs and slugs seen as the difference between
excitability and bistablilty in Eqs.~\eqref{eq:q_pde}-\eqref{eq:u_pde}. Phase
planes show nullclines at (a) $r = 0.7$ and (b) $r = 1$. The fixed point
$(1,0)$ corresponds to parabolic flow. In (b) the additional stable fixed
point corresponds to stable turbulence. Solution snapshots show (c) a puff at
$r = 0.7$ and (d) a slug at $r = 1$. These solutions are plotted in the phase
planes with arrows indicating increasing $x$.  }
\label{fig:pde}
\end{figure}

The core of the model is seen in the $q$-$u$ phase plane in
Fig.~\ref{fig:pde}.
The trajectories are organized by the nullclines: curve where $\dot u = 0$ and
$\dot q=0$ for the local dynamics ($q_{xx} = q_{x} = u_{x} = 0$).
For all $r$ the nullclines intersect in a stable, but excitable, fixed point
corresponding to laminar parabolic flow. 
The $u$ dynamics with $\epsilon_2 > \epsilon_1$ captures in the simplest way
the behavior of the mean shear. In the absence of turbulence ($q=0$), $u$
relaxes to $u=1$ at rate $\epsilon_1$, while in response to turbulence
($q>0$), $u$ decreases at a faster rate dominated by $\epsilon_2$.
Values $\epsilon_1 = 0.04$ and $\epsilon_2 = 0.2$ give reasonable agreement
with pipe flow. (See the Appendix Sec.~\ref{sec:continuous}.)
The $q$-nullcline consists of $q=0$ (turbulence is not spontaneously generated
from laminar flow) together with a parabolic curve whose nose varies with $r$,
while maintaining a fixed intersection with $q=0$ at $u=1+\delta$, ($\delta =
0.1$ is used here).  The upper branch is attractive, while the lower branch is
repelling and sets the nonlinear stability threshold for laminar flow. If
laminar flow is perturbed beyond the threshold (which decreases with $r$
like $r^{-1}$), $q$ is nonlinearly amplified
and $u$ decreases in response.

The (excitable) puff regime occurs for $r < r_c \simeq \epsilon_2/(\epsilon_1
+ \epsilon_2)$, Figs.~\ref{fig:pde}(a) and (c).
The upstream side of a puff is a trigger front ~\cite{TYSON:1988p1143} where
abrupt laminar to turbulent transition takes place. However, turbulence 
cannot be maintained locally following the drop in the mean shear. The system
relaminarizes (reverse transition) on the downstream side in a phase
front~\cite{TYSON:1988p1143} whose speed is set by the upstream front.
Following relaminarization, $u$ relaxes and laminar flow regains
susceptibility to turbulent perturbations.
The slug regime occurs for $r > r_c$, Figs.~\ref{fig:pde}(b) and (d). The
nullclines intersect in additional fixed points.  The system is bistable and
turbulence can be maintained indefinitely in the presence of modified shear.
Both the upstream and downstream sides are trigger fronts, moving at different
speeds, giving rise to an expansion of turbulence. A full analysis will be
presented elsewhere.

\begin{figure}
\centering
\includegraphics[width=3.375in,clip]{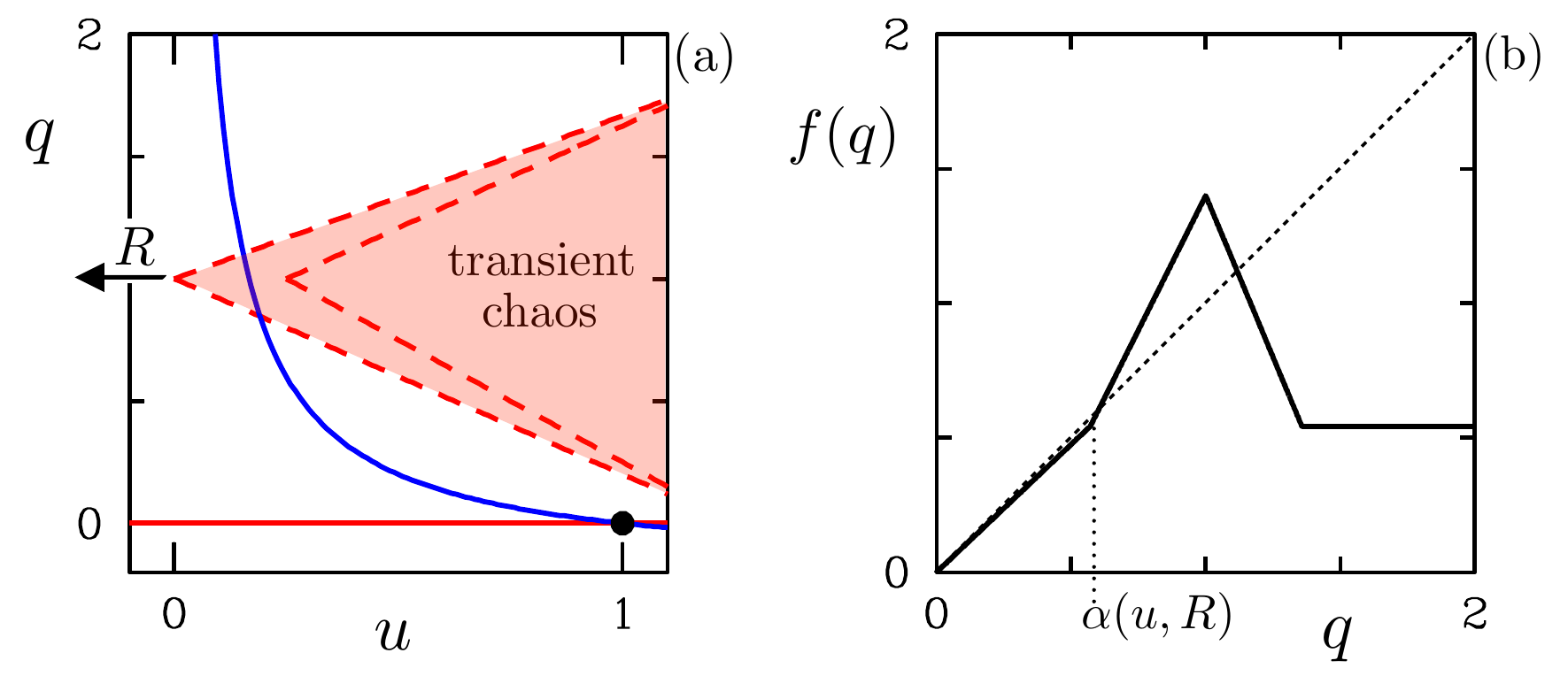}
\caption
{(Color online) Illustration of the discrete model. (a) Local dynamics in the
  $u$-$q$ phase plane. Within a wedge-shaped region $q$ undergoes transient
  chaos, while outside it decays monotonically to $q=0$.  The region varies
  with $R$ as indicated. (b) Map used to produce transient chaos. Parameter
  $\alpha$ (which depends on $u$ and $R$), is the lower boundary separating
  monotonic and chaotic dynamics. }
\label{fig:model}
\end{figure}

While Eqs.~\eqref{eq:q_pde}-\eqref{eq:u_pde} capture the basic properties of
puffs and slugs, the turbulence model is too simplistic to show puff decay and
puff splitting.
Evidence suggests that pipe turbulence is locally a chaotic
repeller~\cite{Eckhardt:2007p887}. Hence a more realistic model, within the
two-variable excitability setting, is obtained by replacing the upper
turbulent branches in Fig.~\ref{fig:pde} with a wedged-shaped region of
transient chaos, illustrated in Fig.~\ref{fig:model}. Outside this region
$q$ decays monotonically.
The model is
\begin{align}
    q_{i+1}^{n+1} & = F(q_i^n + d (q_{i-1}^n - 2 q_i^n + q_{i+1}^n), u_i^n), 
    \label{eq:q_map} \\
    u_{i+1}^{n+1} & = u_i^n + \epsilon_1 (1 - u_i^n) - \epsilon_2 u_i^n q_i^n
        -c ( u_i^n - u_{i-1}^n ),
    \label{eq:u_map} 
\end{align}
where $q_i^n$ and $u_i^n$ denote values at spatial location $i$ and time $n$.
This model is essentially a discrete version of
Eqs.~\eqref{eq:q_pde}-\eqref{eq:u_pde}, except with chaotic $q$ dynamics
generated by the map $F$.

The map $F$ is based on models of chaotic repellers in shear
flows~\cite{Bottin:1998p305, Vollmer:2009p1066}.
Consider the tent map $f$ given by
\begin{align}
    f(q) = 
\begin{cases}
\gamma q                & \mbox{if~ } q < Q_1 \\
2q - \alpha             & \mbox{if~ } Q_1 \le q < 1 \\
4 + \beta - \alpha -(2+\beta)q   & \mbox{if~ } 1 \le q < Q_2 \\
\gamma Q_1            & \mbox{if~ } Q_2 \le q 
\end{cases}
\label{eq:map}
\end{align}
with $Q_1 = \alpha / (2 - \gamma)$ and $Q_2 = (4 + \beta - \alpha - \gamma
Q_1) / (2 + \beta)$.  Parameter $\alpha$ marks the lower boundary separating
chaotic and monotonic dynamics, Fig.~\ref{fig:model}(b), while $\gamma$ sets
the decay rate to the fixed point $q=0$.  For $\beta>0$ ($\beta<0)$ the map
generates transient (persistent) chaos within the tent region.
The map is incorporated into the pipe model by having the threshold $\alpha$
depend on $u$ as well as on a control parameter $R$, via
\begin{equation}
\alpha =  2000 (1 - 0.8 u) R^{-1}.
\label{eq:alpha}
\end{equation}
The factor $(1 - 0.8 u)$ generates the desired wedged-shaped region, while 
2000 sets the scale of $R$ to that of $Re$.
Finally, the map $F$ is given by $k$ iterates of $f$, i.e.\ $F = f^k$; with
$k=2$ used here. (See the Appendix Sec.~\ref{sec:discrete}.)  
This has the effect of increasing the Lyapunov exponent within the chaotic
region.

The only important new parameter introduced in the discrete model is $\beta$
since it quantifies a new effect -- spontaneous decay of local turbulence for
$\beta>0$.
Suitable values for others are: $\epsilon_1 = 0.04$ and $\epsilon_2 = 0.2$ as
before, $\gamma = 0.95$, $c=0.45$ and $d=0.15$. (See the Appendix
Sec.~\ref{sec:discrete}.)
As shown in Fig.~\ref{fig:DNS_model}, for $\beta=0.4$ the model shows
puffs, puff splitting, and slugs remarkably like those from full DNS.  
The model parameter $R$ nearly corresponds to Reynolds number $Re$.

While positive $\beta$ is ultimately of interest, to better connect the two
models consider first $\beta$ negative, e.g.\ $\beta=-0.4$. A transition from
puffs to slugs occurs as $R$ increases and the wedge of chaos crosses the
$u$-nullcline. One finds a noisy version of the continuous model in
Fig.~\ref{fig:pde}.  (See the Appendix Fig.~\ref{fig:compare_betas}.)  If
splittings of turbulent patches occur, they are exceedingly rare. At
$\beta \approx 0$, (including even $\beta = -0.1$), chaotic fluctuations in
$q$ cause occasional splitting of expanding turbulence. Puffs at lower $R$ are
clearly metastable, persisting for long times before decaying.
However, splitting and decay are unrealistically infrequent if $\beta$ is too
small. Setting $\beta \gtrsim 0.1$ gives realistic behavior, as seen in
Fig.~\ref{fig:DNS_model} where $\beta=0.4$.
(See also the Appendix Fig.~\ref{fig:compare_betas_prime}.)
Note that the splitting of expanding turbulent patches and the decay of
localized puffs are caused by the same process - the collective escape from
the chaotic region of a sufficiently large streamwise interval to bring about
local relaminarization. This is precisely the scenario described by extreme
fluctuations~\cite{Goldenfeld:2010}. In the case of puffs, this results in
puff decay, while in the case of splitting, laminar gaps open whose sizes are
then set by the recovery of the slow $u$ field.

\begin{figure}
\centering
\includegraphics[width=3.325in]{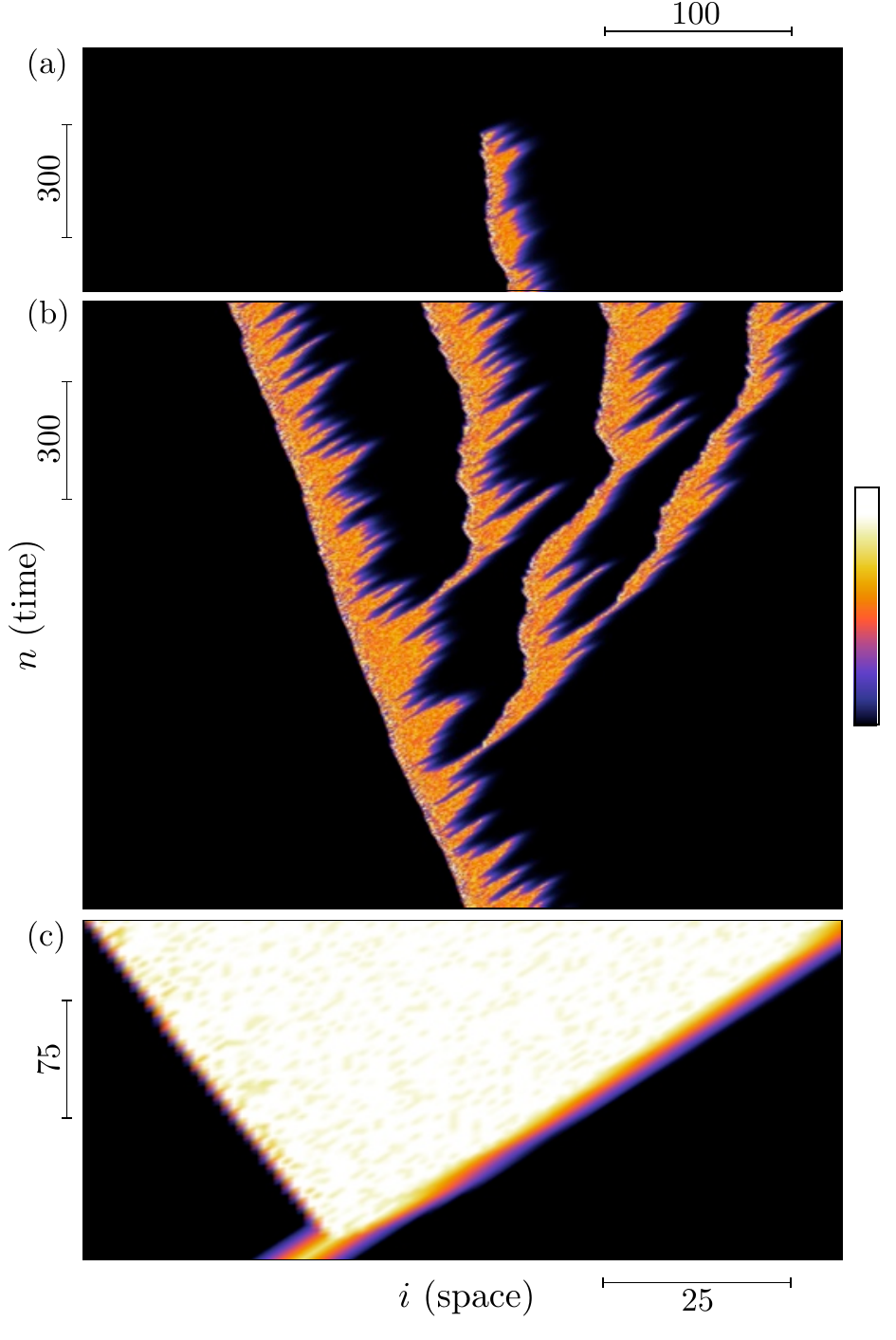}
\caption 
{(Color online) Three regimes of pipe flow from simulations of the discrete
model~\eqref{eq:q_map}-\eqref{eq:alpha}. Space-time diagrams (time upward)
illustrate (a) decaying puff at $R=1900$, (b) puff splitting at $R=2200$, and
(c) slug formation from an edge state at $R=3000$.  For ease of comparison
with published work on pipe flow, solutions are shown in a frame co-moving
with structures. Turbulence intensity $q$ is plotted with $q=1.8$ in white. In
(a) and (b) the scale is linear with $q=0$ in black, while in (c) the scale is
logarithmic with $q \le 10^{-3}$ in black.  Dimension bars indicate space and
time scales. The top space scale applies also to (b).  }
\label{fig:visu}
\end{figure}

Figure~\ref{fig:visu} further illustrates how well the discrete model captures
the three regimes of transitional pipe flow. Spacetime plots show puff decay,
puff splitting, and slug flow.  In Fig.~\ref{fig:visu}(a), a puff persist for
only a finite time before abruptly decaying~\cite{Faisst:2004p898,
Peixinho:2006, Hof:2006p607, Willis:2007p833, Schneider:2008p642,
Hof:2008p914, Avila:2010p839, Kuik:2010p911}.  In Fig.~\ref{fig:visu}(b), puff
splitting dominates the dynamics~\cite{Wygnanski:1975, Nishi:2008, Moxey:2010,
Avila:2011}. New puffs are generated downstream from existing ones such that
intermittent turbulent regions fill space. Compare especially with
Refs.~\onlinecite{Moxey:2010, Avila:2011}.  Finally, in
Fig.~\ref{fig:visu}(c), a slug arises from a localized edge state (a
low-amplitude state on the boundary separating initial conditions which evolve
to turbulence from those which decay to laminar flow~\cite{Schneider:2007p825,
Schneider:2009p54, Mellibovsky:2009, Duguet:2010p2}). Compare especially with
Ref.~\onlinecite{Duguet:2010p2}.


\begin{figure}
\centering
\includegraphics[width=3.375in]{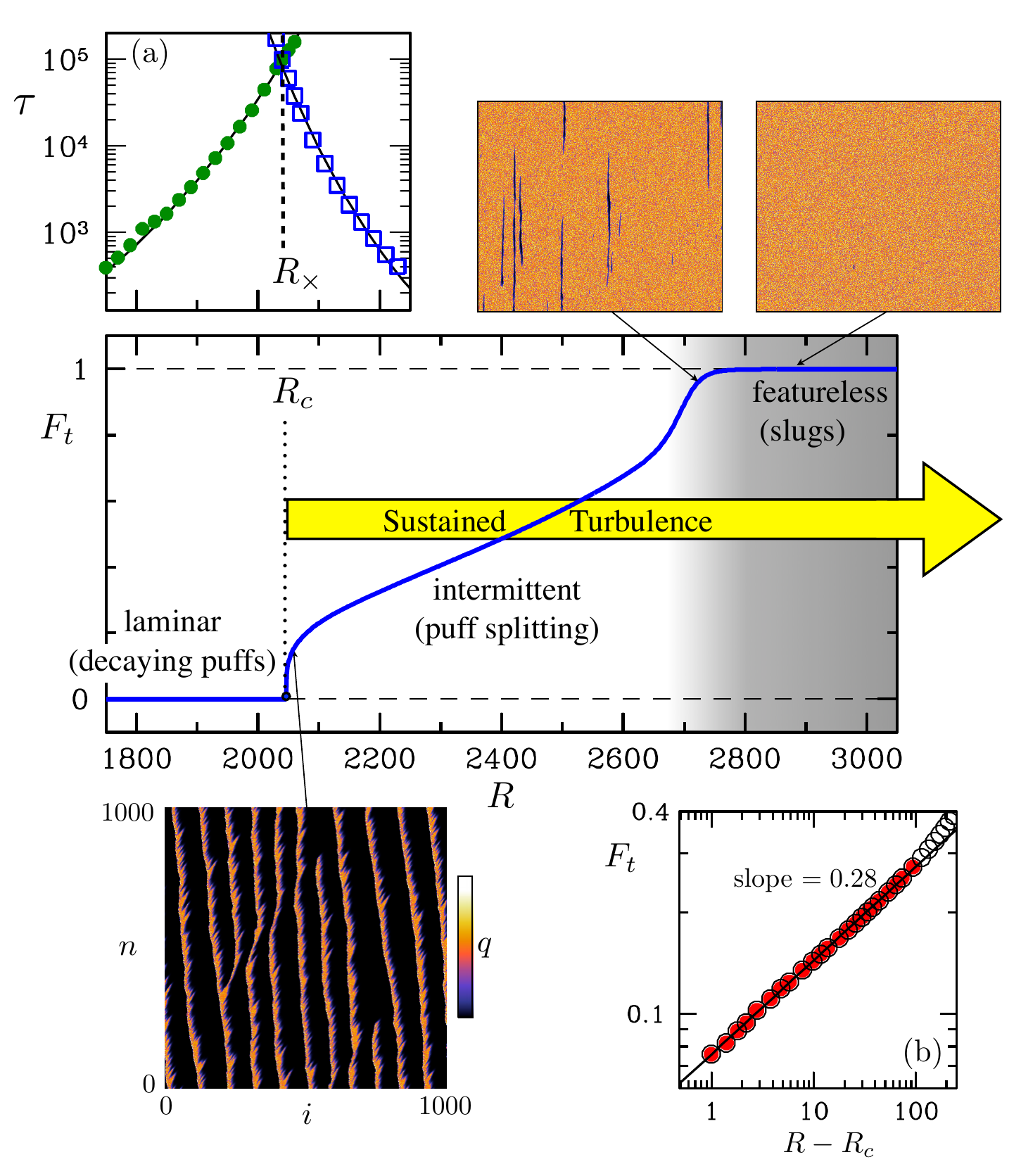}
\caption
{(Color online) Main figure is a bifurcation diagram for model turbulence in
the thermodynamic limit.  The turbulence fraction $F_t$ is plotted throughout
the transitional regime.
The onset of sustained turbulence, via spatiotemporal
intermittency, occurs continuously at
$R_c \simeq 2046.2$. $F_t$ increases with $R$ and saturates near $R = 2800$.
Asymptotic regimes (laminar, intermittent, featureless) are labeled, along
with corresponding transient dynamics (decaying puffs, puff splitting,
slugs). The onset of featureless turbulence is not sharp as indicated by gray
shading.
Space-time plots illustrate the dynamics near the ends of the transitional
regime ($R = 2058$, $R = 2720$, $R = 2880$) with $q$ plotted in frames
co-moving with structures (color map indicated, dark is laminar).
(a) Mean lifetimes for decaying (circles) and splitting (squares) puffs
crossing at $R_\times \simeq 2040$.
(b) log-log plot of $F_t$ versus $R - R_c$.  Best fit to the filled (red)
points determines $R_c$ and the slope.  
}
\label{fig:Ft}
\end{figure}

The remainder of the paper provides a global perspective of the
transitional regime, obtained from extensive numerical simulation of
Eqs.~\eqref{eq:q_map}-\eqref{eq:u_map} and summarized in Fig.~\ref{fig:Ft}.
Turbulence fraction $F_t$ serves as the order parameter, 
tracking the dynamics from 
the onset of intermittency through
the approach to uniform, featureless turbulence.
A point is defined to be turbulent if $q > 0.5 \alpha$ and $F_t$ is the mean
fraction of turbulent points. 


There is a continuous transition to sustained turbulence, via spatiotemporal
intermittency~\cite{Pomeau:1986, Chate:1988p8, Chate:1988p48, Bottin:1998p305,
Manneville:2009p316, Avila:2011}, at a critical value $R_c \simeq
2046.2$. Below $R_c$, the flow is asymptotically laminar and $F_t = 0$. Above
$R_c$ turbulence persists indefinitely and $F_t > 0$.
This transition is associated with the crossing of mean lifetimes for puff
decay and splitting shown in Fig.~\ref{fig:Ft}(a).  Both decay and splitting
are memoryless processes with exponential survival distributions
$P \sim \exp(-n/\tau(R))$, where $\tau(R)$ is the $R$-dependent mean time
until decay or split.  
(See the Appendix Sec.~\ref{sec:stats} and Fig.~\ref{fig:lifetimes}.)
The mean lifetimes vary approximately
super-exponentially with $R$ ~\cite{Hof:2008p914, Avila:2010p839,Avila:2011},
but neither is exactly of this form.  Mean lifetimes cross at
$R_\times \simeq 2040$.  Above $R_\times$ an isolated puff is more likely to
split than decay.  
As expected, even though individual turbulent domains may still decay, others
may split, as seen in the spacetime plot at $R = 2058$.
Due to correlations between splitting and decay events, the
critical value $R_c$ is not identical to $R_\times$, but is very close to it
(a difference of $0.3\%$).
Fig.~\ref{fig:Ft}(b) shows that just above criticality, $F_t \sim (R -
R_c)^{0.28}$, supporting that the transition falls into the universality class
of directed percolation ~\cite{Hinrichsen:2000}.

The ratio of turbulence to laminar flow increases through the intermittent
region and at the upper end only small laminar flashes are seen within a
turbulent background. Beyond $R \simeq 2800$ laminar regions essentially
disappear and $F_t \simeq 1$. This occurs in pipe flow at $Re \simeq
2600$~\cite{Moxey:2010}.
The transition to featureless turbulence is not sharp, however, nor is the
transition from puff splitting to slugs.
This upper transition will be addressed elsewhere, 
but the
basic effect, common for bistable media, is already contained in
Eqs.~\eqref{eq:q_pde}-\eqref{eq:u_pde}. For a range of $r$ above the slug
transition, ($r_c < r \lesssim 0.91$), 
turbulence does not expand to fill the domain in the presence of other
slugs. Small laminar regions remain due to the recovery of the slow $u$-field
and this sets the scale for the laminar flashes at the upper end of the
transition region in Fig.~\ref{fig:Ft}.


I have sought to understand key elements of transitional pipe flow --
puffs, puff splitting, and slugs -- without appealing in detail to the 
underlying structures within shear turbulence.
This approach is similar to that expounded by Pomeau~\cite{Pomeau:1986}, and
considered elsewhere~\cite{Bottin:1998p305}.
The important insight here is the close connection between subcritical shear
flows and excitable systems.
The view is that a great many features of intermittent pipe flow can be
understood as a generic consequence of the transition from excitability to
bistability where the turbulent branch is itself locally a chaotic repeller.
I have introduced particular model equations to express these ideas in simple
form. While phenomena have been demonstrated with specific parameters,
the phenomena are robust.
The challenge for future work is to obtain more quantitatively accurate
models, perhaps utilizing full simulations of pipe flow, since ultimately the
fluid mechanics of shear turbulence (streaks and streamwise vortices) is
important for the details of the process.
More challenging is to extend this effort to other subcritical shear flows,
such a plane channel, plane Couette, and boundary-layer flows. These 
require non-trivial
extensions of the current work because, unlike here, the mean shear profile
cannot obviously be well captured by a simple scalar field.

\begin{acknowledgments}
  I thank M.\ Avila, Y.\ Duguet, B.\ Hof, P.\ Manneville, D.\ Moxey, and L.\
  Tuckerman for valuable discussions. Computing resources were provided by
  IDRIS (grant 2010-1119).
\end{acknowledgments}

\appendix*
\section{Supplemental Information}
\label{sec:appendix}

\subsection{Parameters Selection}

Here the parameter selection used in this study is discussed. No attempt has
been made to determine precisely values such for the best fit to pipe flow.
The models are not sufficiently quantitative that exact comparisons are called
for at this time.  Moreover, the phenomena presented in the paper are very
robust and for some parameters there simply is not a strong criterion to use
to select precise values.  The goal is to provide justification for the values
used in the paper as well as insight into how the parameters control the
dynamics of the models.

\subsubsection{Parameters for continuous model}
\label{sec:continuous}

\begin{figure}[h]
\centering
\includegraphics[width=3.0in]{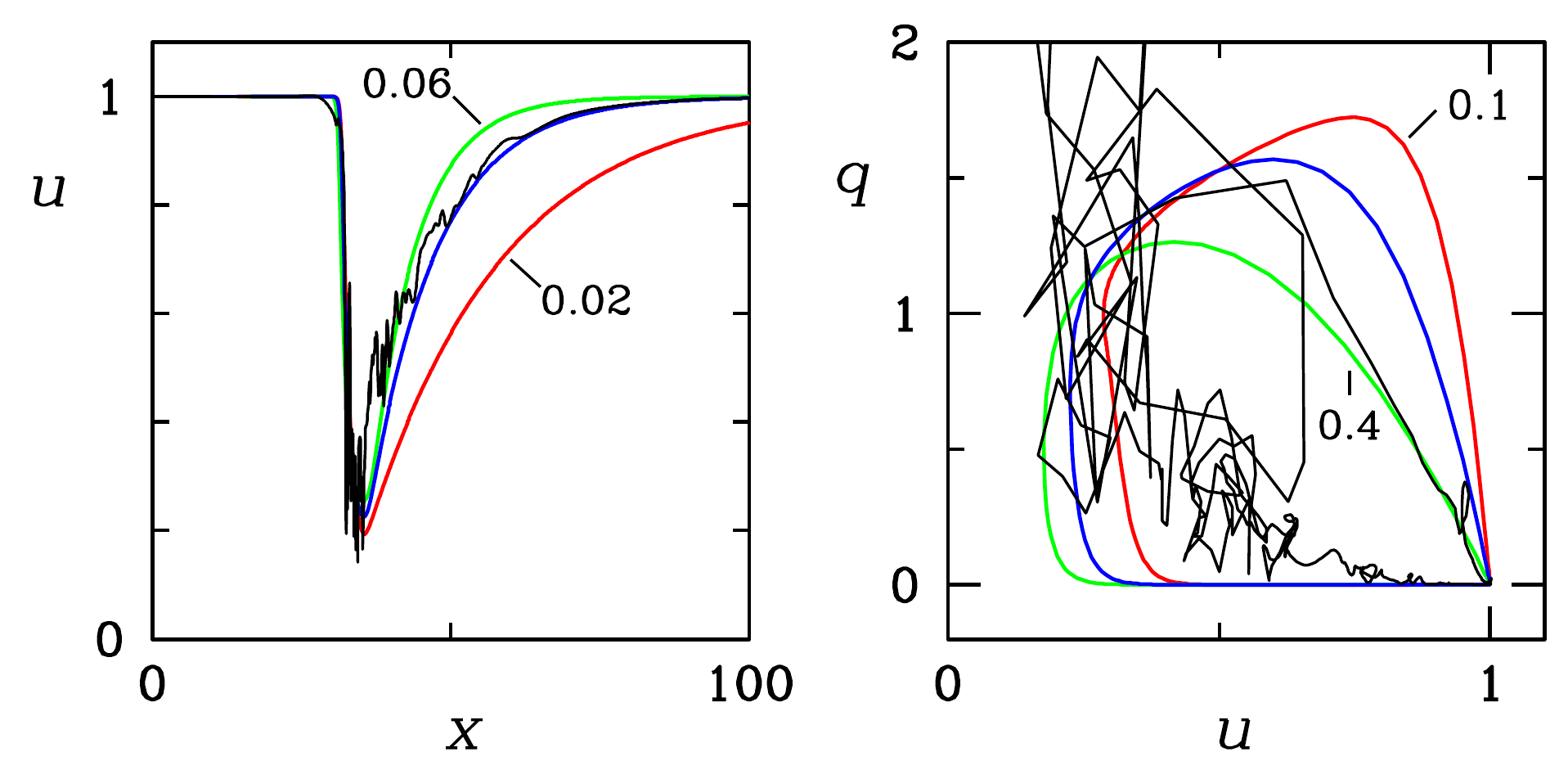}
\caption
{(Color online) 
Parameters $\epsilon_1$ and $\epsilon_2$ chosen to match a typical puff from
DNS at $Re=2000$. In the left plot $\epsilon_2 = 0.2$ while $\epsilon_1$ has
values 0.02 (red), 0.04 (blue), and 0.06 (green). In the right plot
$\epsilon_1 = 0.04$ and $\epsilon_2$ has values 0.1 (red), 0.2 (blue), and 0.4
(green). $r=0.7$. DNS is the irregular black curve.  The unlabeled (blue)
curves in the two plots correspond to the values of $\epsilon_1$ and
$\epsilon_2$ used in the paper. }
\label{fig:compare_epsilons}
\end{figure}

Only the two rates $\epsilon_1$ and $\epsilon_2$ need to be determined. The
value of $\delta$ has little impact on the dynamics and has simply been fixed
at $0.1$. These parameters are determined by fitting to a typical puff from
DNS as shown in Fig.~\ref{fig:compare_epsilons}.

The left plot shows the spatial profile of model puffs for three values of
$\epsilon_1$, the parameter controlling the final relaxation to parabolic
flow. It is straightforward to select a reasonable value of $\epsilon_1$ from
such a plot. Note, however, that a scaling of model length scale has been
performed to plot model and DNS profiles on the same graph (model lengths have
been multiplied by 0.225). This scaling of length is such that the sharp
upstream edge of the model puff occurs over the same distance as in DNS.  The
upstream edge is largely set by $\epsilon_2$. (If the scaling of space units
between model and DNS where known for other reasons, then the spatial profile
alone could be used to determine both $\epsilon_1$ and $\epsilon_2$.) 

The right plot is used then to complete the determination. Here model puffs
for different values of $\epsilon_2$ are plotted in the $u$-$q$ plane. The
sharp upstream edge of a puff is the trajectory rising from parabolic flow at
$u=1$, $q=0$ and this is strongly affected by the value of $\epsilon_2$. If
$\epsilon_2$ is too small then the trajectory is too step ($u$ does not respond
quickly enough). If $\epsilon_2$ is too large, then $q$ does not reach a
sufficiently large value.

The values of $\epsilon_1 = 0.04$ and $\epsilon_2 = 0.2$ chosen for the
simulations presented in the paper were arrived at by varying the two values
to get the best overall agreement in the spatial profile and phase portrait.

\subsubsection{Parameters for discrete model}
\label{sec:discrete}

As stated in the paper, the two rates $\epsilon_1$ and $\epsilon_2$ are taken
to have the same values as in the continuous model. This is quite reasonable
given the relationship between Eqs.~(2) and (4).  This leaves choosing the
parameters $k$, $\beta$, and $\gamma$ of the map $F$, and the parameters $c$
and $d$. The role of each of these is discussed below. As each parameter is
varied in the following, the remaining parameters take the fixed values used
in the paper: $\epsilon_1 = 0.04$, $\epsilon_2 = 0.2$, $k=2$, $\beta=0.4$,
$\gamma=0.95$, $c=0.45$, and $d=0.15$.

\begin{figure}[h]
\centering
\includegraphics[width=3.375in]{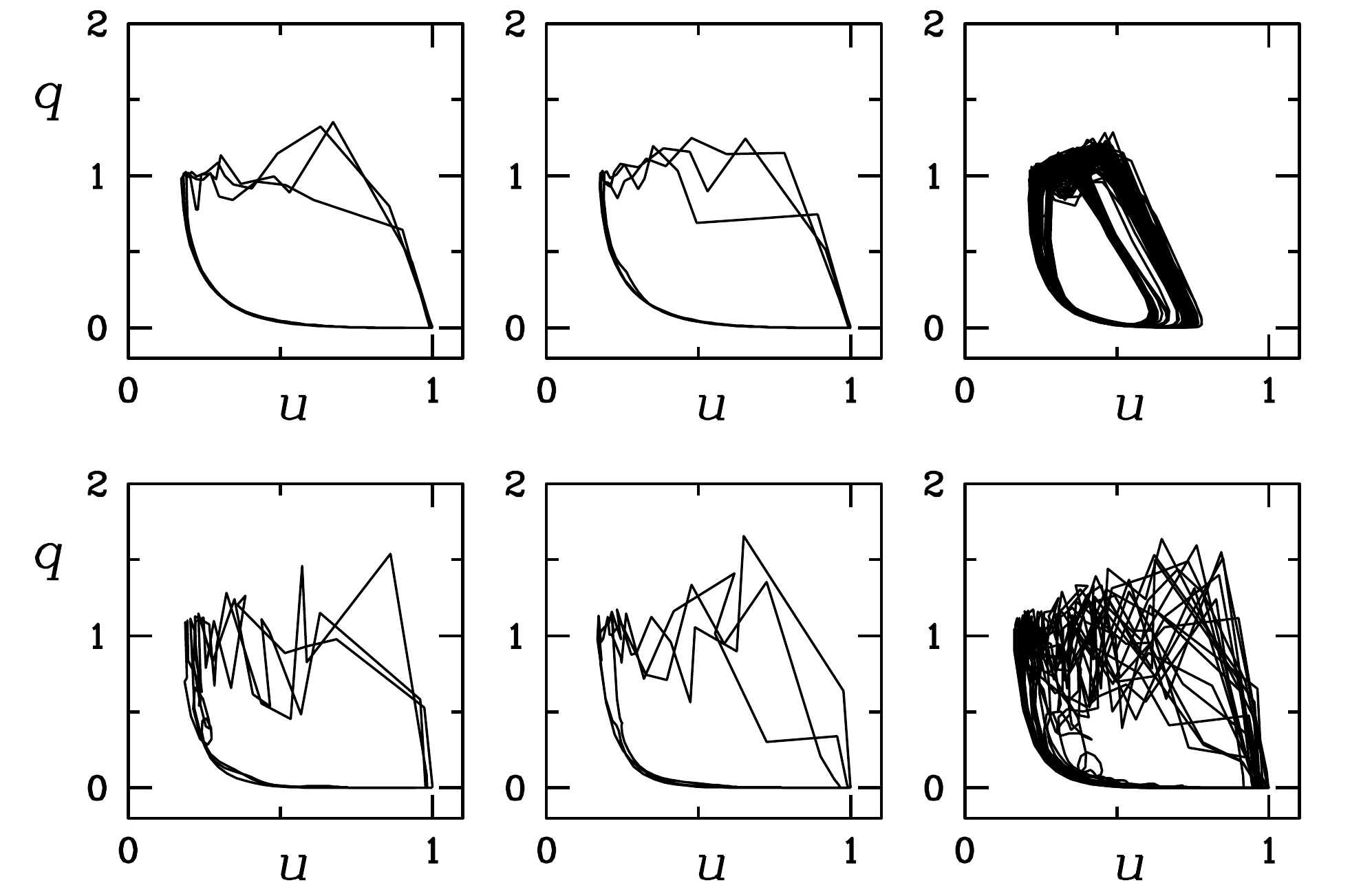}
\caption
{Effect of parameter $k$. Top row $k=1$ and bottom row $k=2$. In each case
snapshots are plotted in the $u$-$q$ phase plane. The left two plots show
randomly chosen snapshots of solutions with three closely spaced puffs. The
right-most plots are of solutions following a quench from high $R$ and contain
a large number of puffs. Top row is at $R=1760$ and the bottom row is at
$R=2000$. Both values of $R$ are close to the transition to sustained
turbulence for the corresponding value of $k$.
}
\label{fig:compare_ks}
\end{figure}

{\it Parameter $k$}:
The parameter $k$ effectively dictates how many iterates of the map $f$ are
used per time step of the model. The effect of the parameter $k$ is shown in
Fig.~\ref{fig:compare_ks} where puff solutions are shown in the $u$-$q$ phase
plane for $k=1$ and $k=2$. Model turbulence is more erratic for $k=2$ than 
$k=1$. 
When $k=1$, puff splitting and the transition to sustained turbulence occurs
at a smaller value of $R$, but the fact that $R$ is smaller on the top row of
Fig.~\ref{fig:compare_ks} only partially accounts for the difference between
the top and bottom rows of Fig.~\ref{fig:compare_ks}.

In addition to this visual comparison, there is the fact that the average
slope for a unimodal map with stable chaos is limited to $\lambda=2$ and this
is an artificial constraint on the dynamics that comes about from considering
one-dimensional dynamics. (In the case of transient chaos the mean slope can
exceed 2, but there is still a constraint relating the escape rate to the
mean slope.)  Taking $k>1$ is equivalent to considering multimodal maps and
removes the artificial constraint.

Note that the model shows puffs, puff splitting and slugs even for
$k=1$. These features are robust. However, the additional freedom in the
chaotic dynamics by allowing $k=2$ permits the model to achieve a better
representation of turbulent flow. I have not found that using $k>2$ 
offers noticeable further improvement.

\begin{figure}
\centering
\includegraphics[width=3.375in]{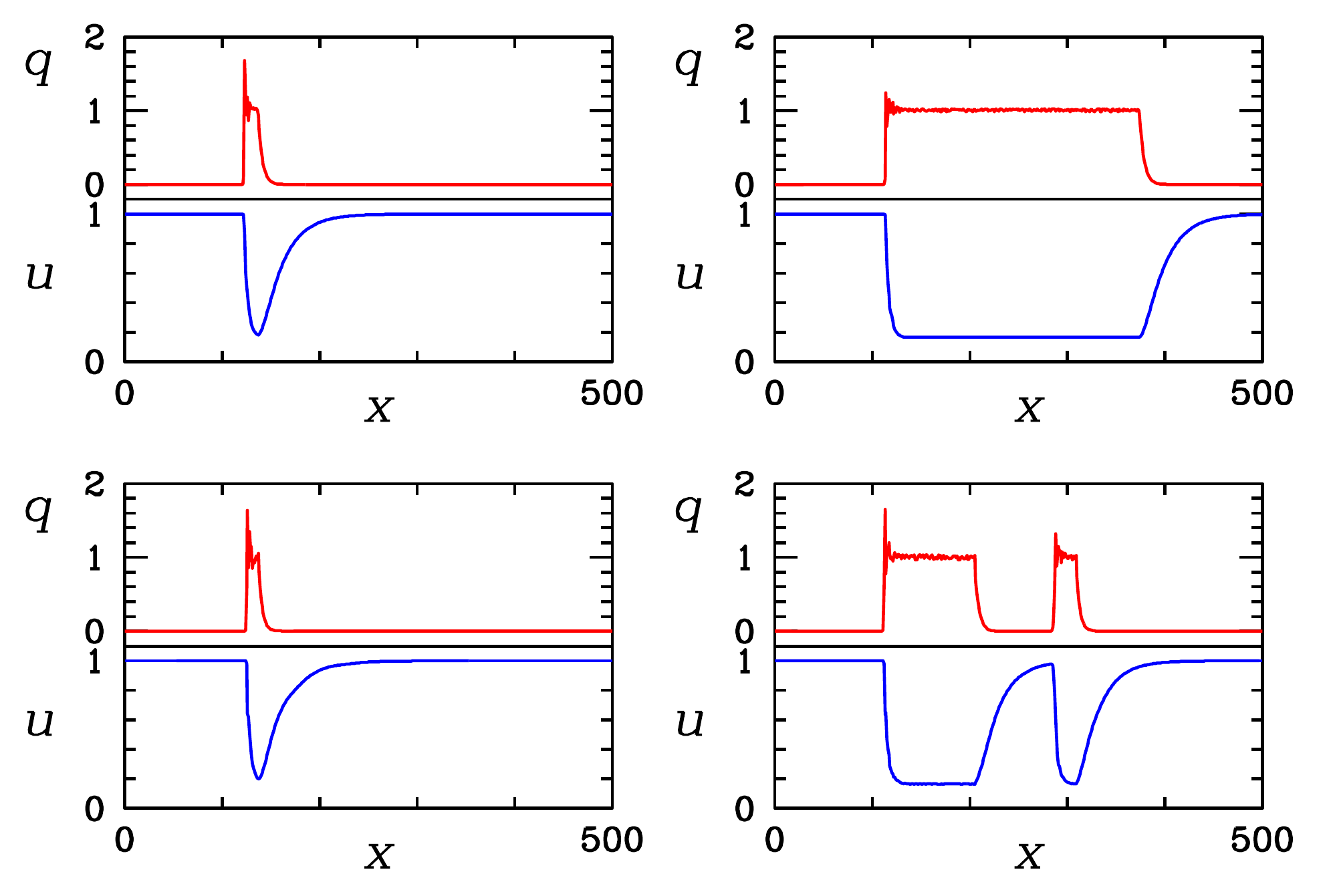}
\caption
{ (color online) 
Effect of parameter $\beta$. Top row $\beta=-0.4$ and bottom row
$\beta=0$. In each case snapshots of solutions are shown on either side on the
transition from localized puffs to expanding turbulence. 
Upper left: $R=1750$. Upper right: $R=1780$. 
Lower left: $R=1760$. Lower right: $R=1800$. 
}
\label{fig:compare_betas}
\end{figure}

{\it Parameter $\beta$}: This parameter controls the leakage rate from the
chaotic region of the map. Figure~\ref{fig:compare_betas} shows examples of
states for $\beta=-0.4$ and $\beta=0$. With $\beta$ sufficiently negative, as
for $\beta=-0.4$ in the top row of Fig.~\ref{fig:compare_betas}, a transition
from puffs to slugs occurs that is essentially just a noisy version of the
continuous models shown in Fig.~2 of the paper. Note that the chaotic wedge
first touches the $u$-nullcline at $R=1733$ and the transition from puffs to
slugs occurs near to, but not exactly at, this value. If there are any
splitting events they are very rare.

With $\beta \approx 0$, as for $\beta=0$ in the bottom row of
Fig.~\ref{fig:compare_betas}, the transition from puffs to slugs is mediated
by splitting events. However, the splitting events are too rare for the model
to realistically correspond to pipe flow.

\begin{figure}
\centering
\includegraphics[width=3.375in]{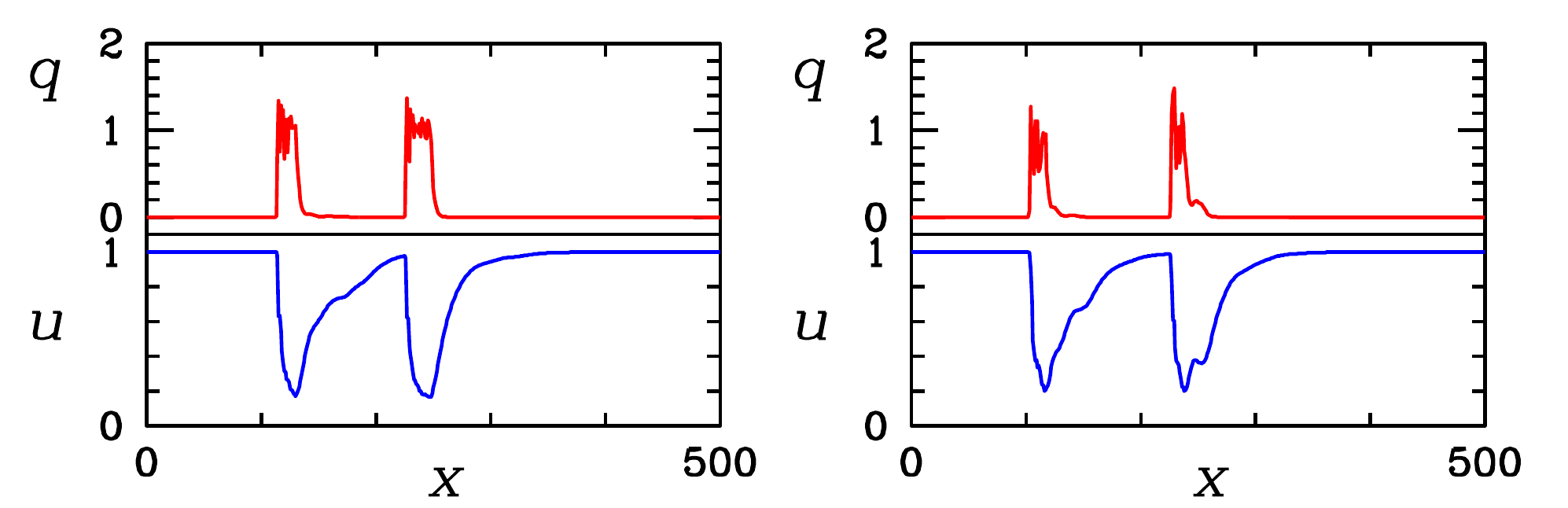}
\caption
{ (color online) 
Further effect of parameter $\beta$. Left $\beta=0.3$ ($R=1980$) and right
$\beta=0.5$ ($R=2180$). In each case snapshots of solutions are shown just
after a puff split. There is little to distinguish the cases. 
}
\label{fig:compare_betas_prime}
\end{figure}

As illustrated in Fig.~\ref{fig:compare_betas_prime}, with $\beta \gtrsim
0.1$, model puff splitting is similar to pipe flow. There appears to be no
strong basis to select any particular value of $\beta$ based on a visual
examination of the onset of splitting. The value $\beta = 0.4$ used in the
paper is simply a representative value.

\begin{figure}
\centering
\includegraphics[width=3.3755in]{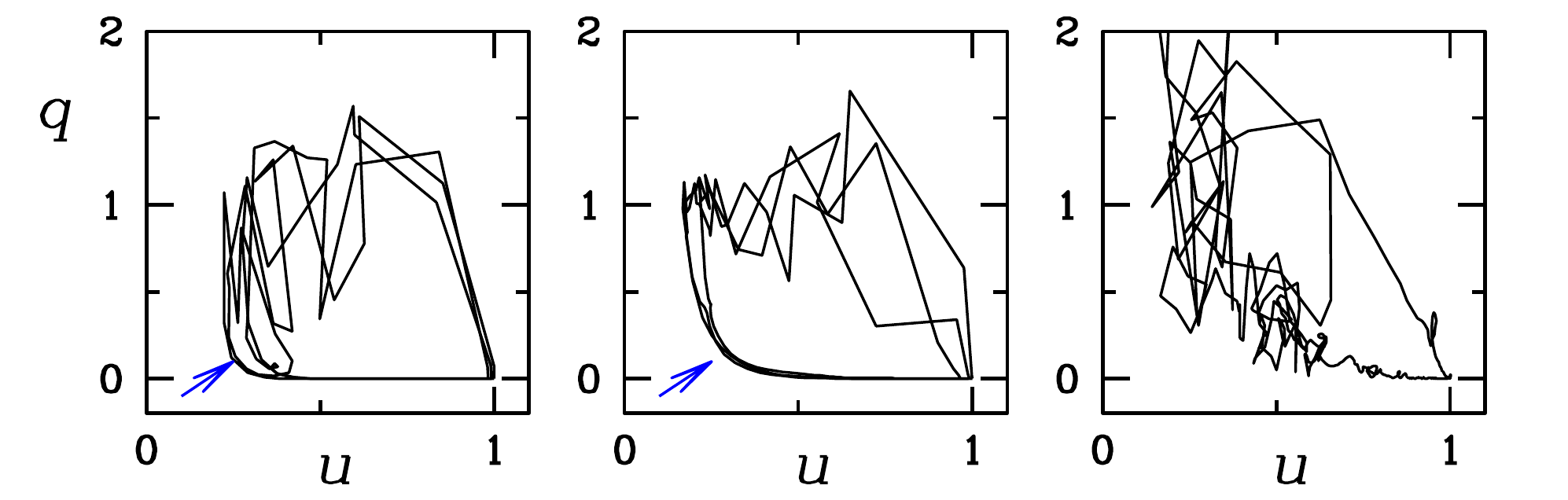}
\caption
{ (Color online) 
Effect of parameter $\gamma$. Left shows $\gamma = 0.80$, with $R=2400$,
middle shows $\gamma = 0.95$, with $R=2000$, and is the same plot as bottom
middle of Fig.\ 2. Arrows indicate the relevant region of the $u$-$q$ phase
plane.
The right-most plot is DNS of a puff at $Re=2000$, exactly the same as in
Fig.\ 1.
}
\label{fig:compare_gammas}
\end{figure}

{\it Parameter $\gamma$}: This parameter controls the monotone decay of
turbulence $q$ following exit from the chaotic
region. Figure~\ref{fig:compare_gammas} shows puffs plotted in the $u$-$q$
phase plane for two value of $\gamma$. For comparison the puff at $Re=2000$
from Fig.~\ref{fig:compare_epsilons} is repeated here. The smaller $\gamma$,
the more quickly $q$ decays and the less rounded the phase-space dynamics is
in the lower left corner of the phase portraits. The best match of turbulent
decay in the model will be at the largest possible value for
$\gamma$. However, as $\gamma$ approaches 1 laminar flow becomes marginally
stable to $q$ perturbations and this is clearly unphysical. The value
$\gamma=0.95$ was chosen as a compromise between the competing requirements of
having $\gamma$ large but not too close to 1.
An improvement could be likely be obtained by having $\gamma$ be a function of
$u$ and $R$, but this introduces additional fitting parameters and is not done
here. 

{\it Parameters $c$ and $d$}: These parameters are naturally thought of as
arising from the discretization of the terms $u_x$ and $q_{xx}$ in the
continuous model.
At present I am not aware of any compelling reasons to select $c$ and $d$ to
particular values other than that they should be small ($d$ must satisfy $d 
\lesssim 1/2$ for stability reasons). 
The value of $c$ was chosen to be less slightly less than 1/2.
The value of $d$ is then to be fixed. Based on the analogy with the continuous
model, and the value chosen for $c$, it could be taken to be $d = c^2$. This
is because discretizing the continuous model with a time step of $\triangle t
= 1$ means that $c = 1/\triangle x$, from an upwind discretization of the
advection term. Then $d$ will be $1/(\triangle x)^2$ from discretization of
the diffusion term. This would give $d = 0.45^2 = 0.2025$. However, adjusting
$d$ downwards from this value to $d=0.15$ places the transition to sustained
turbulence in the model at the critical Reynolds number for pipe flow.  Having
the model match this transition point seems preferable to setting it to the
particular value $0.2025$. Moreover, it is common to vary the diffusive
coupling constant in studies of coupled-map lattices.

\begin{figure}
\centering
\includegraphics[width=3.3755in]{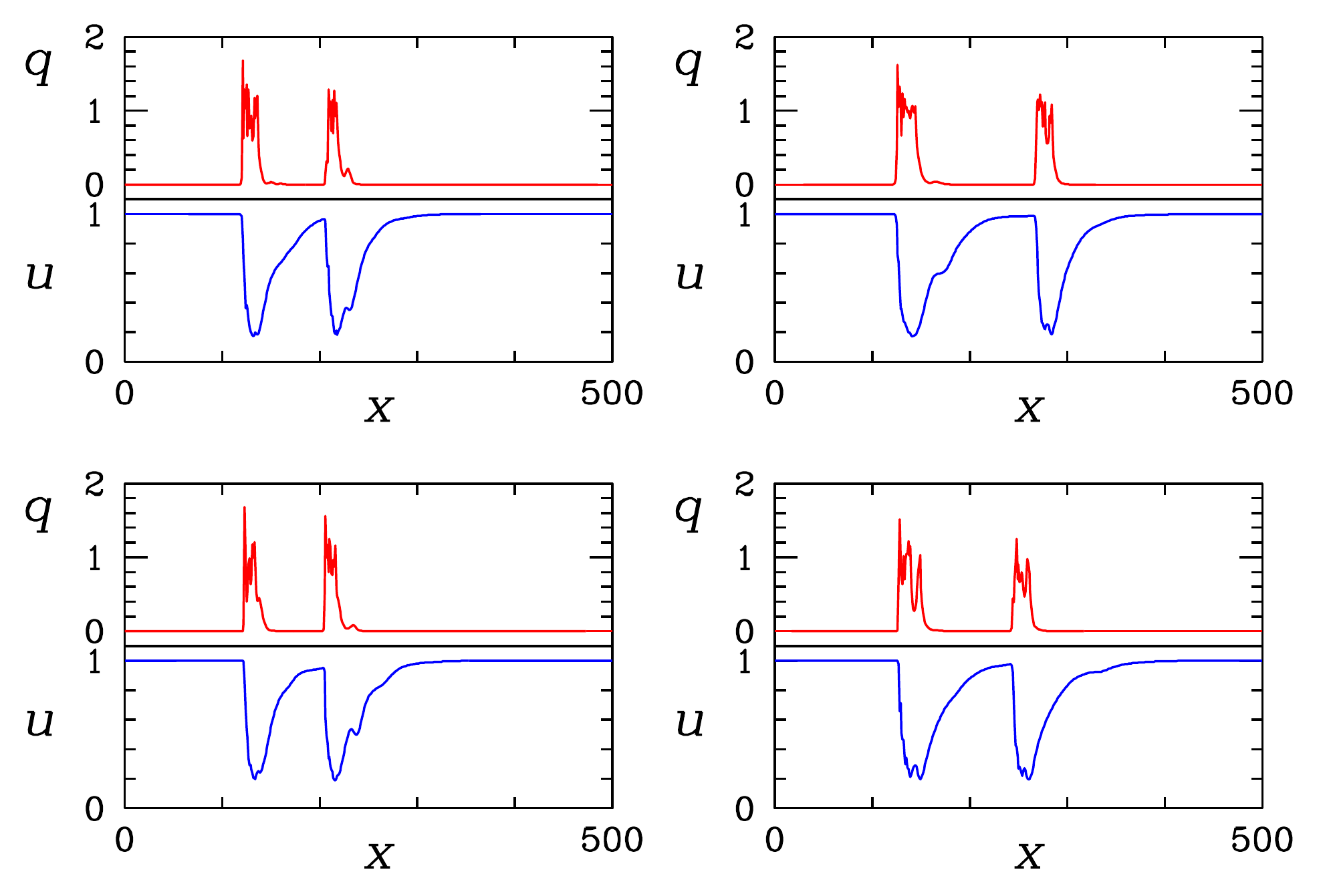}
\caption
{ (Color online) 
Effect of parameters $c$ and $d$.  In each case snapshots of solutions are
shown just after a puff splitting. Splitting is not very sensitive to $c$ and
$d$ around values used in the paper.
Upper left: $c=0.45$, $d=0.1$, $R=2480$. 
Upper right: $c=0.45$, $d=0.2025$, $R=1960$. 
Lower left: $c=0.35$, $d=0.15$, $R=2120$. 
Lower right: $c=0.55$, $d=0.15$, $R=2080$. 
}
\label{fig:compare_cd}
\end{figure}

To emphasize that qualitative features of the model do not strongly dependent
on the parameters $c$ and $d$ (as long as they are reasonably small),
Fig.~\ref{fig:compare_cd} shows some splitting puffs for different values of
$c$ and $d$. The reason for focusing on splitting puffs is these best show the
fidelity of the model. Puffs and slugs are easily obtained. In each case a
puff was generated and $R$ was increased slowly until a splitting occurred. The
value of $R$ are given in the caption.

\subsection{Details of Numerical Study in Figure 4}

\subsubsection{Decay and Splitting Statistics}
\label{sec:stats}


\begin{figure}
\centering
\includegraphics[width=3.375in]{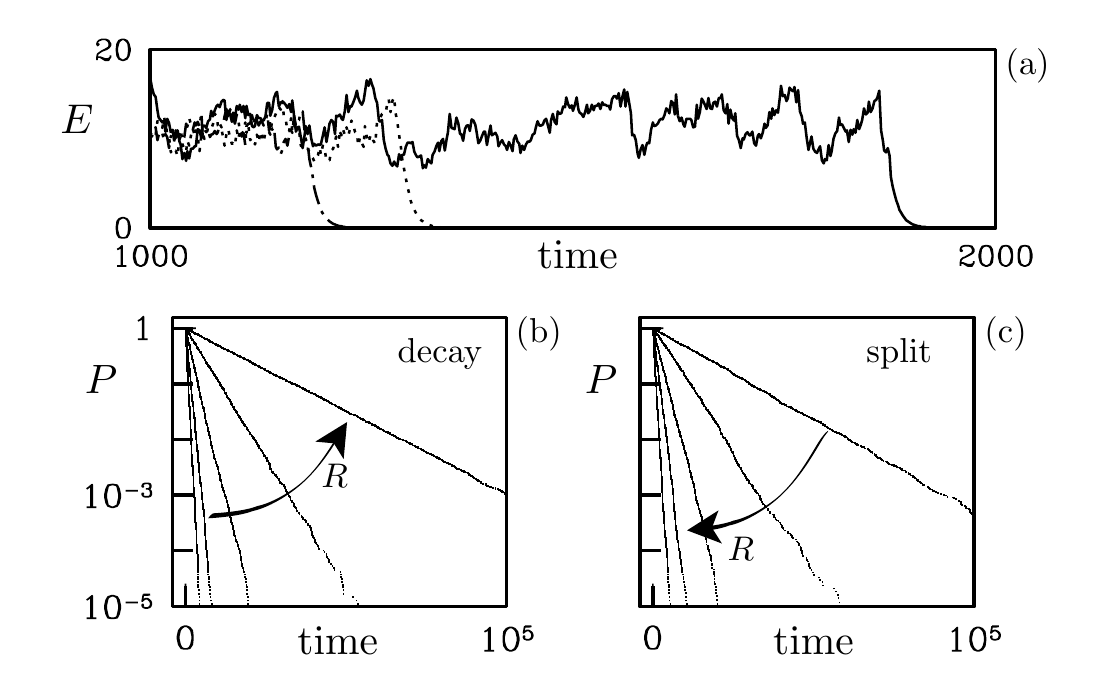}
\caption
{ 
Lifetime statistics for decaying and splitting model puffs. (a) Time series of
total energy $E$ for three puffs illustrating abrupt decay at unpredictable
times ($R=1800$). Bottom row shows exponential (memoryless) probabilities $P$
for (b) decaying puffs ($R=1800, 1850, 1900, 1950, 2000$) and (c) splitting
puffs ($R =2060, 2090, 2120, 2150, 2180$).
}
\label{fig:lifetimes}
\end{figure}

Figure~\ref{fig:lifetimes}(a) shows time series of total energy $E = \sum_i
q_i$ for puffs from three initial conditions in which $q$ is varied at one
space point by less than $10^{-5}$. While the dynamics are deterministic,
abrupt decay occurs at unpredictable times.
From a large number of such simulations lifetime statistics can be generated.
Specifically, simulations were initiated from small perturbations designed to
trigger puffs. A random number generator was used to introduce a small amount
of randomness into each perturbation.  Simulations were then run for 1000 time
steps to allow puffs to equilibrate. If puffs had not decayed, the puffs were
used as initial conditions for simulations for decay statistics.
Figure~\ref{fig:lifetimes}(b) shows representative survival probabilities $P$
of a puff lasting at least time $n$. Each distribution corresponds to 4000
realizations.
The survival functions are exponential (memoryless),
$P \sim \exp(-n/\tau(R))$, where $\tau(R)$ is the $R$-dependent mean
lifetime until decay.

Puff splitting is similar. Initial conditions were generated in the same way
only here equilibration simulations were run for 1400 time steps because at
the largest $R$ in the study 1000 time steps was not quite enough to remove
all equilibration effects.  A puff was defined to have split once two
turbulent peaks are separated by least 80 grid points.
Figure~\ref{fig:lifetimes}(c) shows representative survival probabilities $P$
for a puff to last at least time $n$ without splitting.  The distributions are
again exponential, $P \sim \exp(-n/\tau(R))$, showing that model splitting is
indeed memoryless with a mean splitting time $\tau(R)$.
These lifetimes are plotted in Fig.\ 4(a) of the paper. 

\subsubsection{Turbulence Fraction}

The turbulence fraction $F_t$ serves as the order parameter for the onset of
sustained turbulence.
A point is defined to be turbulent if $q > \kappa \alpha$ where $\kappa$ sets
a threshold relative to the lower boundary separating chaotic and monotonic
decay of the map $F$.
$F_t$ is the mean fraction of grid points in the turbulent state. Means have
been computed from 4 independent simulations. For $R$ near the critical point,
simulations of $8 \times 10^6$ time steps were run on grids of $12 \times
10^4$ points. The standard deviation from the 4 independent simulations is
comparable to the point size in Fig.\ 4(b) of the paper.
For $R>2200$ simulations for $10^6$ time steps on grids of $10^4$ points were
more than sufficient.
In Fig.\ 4 of the paper, $F_t$ is plotted for $\kappa = 0.5$.  While the exact
value of $F_t$ depends on threshold $\kappa$, the extent of the intermittent
region and critical scaling at onset do not.


%

\end{document}